\documentclass[showpacs,preprintnumbers,superscriptaddress,amsmath,amssymb]{revtex4}
\usepackage{amsfonts}
\usepackage{amsmath}
\usepackage{color}
\usepackage{amssymb}
\usepackage{graphicx}
\usepackage{epsfig}
\usepackage{mathrsfs}
\begin{document}
\title{Excluding the local hidden variable theory with time-reversal Bell test}

\author{Yong-gang Tan}
\email{ygtan@lynu.edu.cn} \affiliation{Physics and Information
Engineering Department, Luoyang Normal College, Luoyang 471022,
Henan, People's Republic of China}

\begin{abstract}

A time-reversal Bell test protocol is proposed. The quantum states
are prepared by faraway separated partners and transferred to the
third partner who carries out Bell basis measurement on them to
post-select the Einstein-Podolsky-Rosen (EPR) pairs. If some
loopholes open, similar as that in normal Bell test, the Bell
violation in the present protocol is apt to be interpreted with
local hidden variable (lhv) theory. With some modifications on the
protocol, the lhvs at both sides are prevented to exchange their
information. Thus they only function locally and cannot affect the
behaviors of the states at the other sides. However, Bell violation
can still be obtained in this case. It means that Bell violation is
realized with the lhv theory excluded. Because high detection
efficiency is not compulsory, this protocol can be realized with
present technology.


\end{abstract}
\maketitle

\section{Introduction}

In 1935, Einstein, Podolsky, and Rosen argued that the wavefunction
description of quantum physics is incomplete unless lhv
exists~\cite{Einstein35}. In 1964, Bell proposed the first
experimentally realizable model to test whether lhv exists or
not~\cite{Bell65}. Presently, the most widely used Bell-type
inequalities are the Clauser-Horne-Shimony-Holt (CHSH)
inequality~\cite{Clauser69} and the Clauser-Horne (CH)
inequality~\cite{Clauser74}. Unfortunately, there is no conclusive
verdict on the correctness of the EPR's standpoint due to the
loopholes existing in current Bell experiments
\cite{Freedman72,Aspect81,Aspect82a,Aspect82b,Ou88,Shih88,Kwiat95,
Weihs98,Tittel98,Rowe01,Ansmann09,Hofmann12,Scheidl10}. Therefore,
the realization of loophole-free Bell experiments becomes one of the
most significant issues in fundamental physics.

Normally, the locality loophole exists in Bell test with massive
particles, while the detection loophole is often present in Bell
test with optical system. For Bell test carried out with optical
systems, it has been shown that the detection loophole can be closed
only when the detection efficiency $\eta$ is higher than
$82.8\%$~\cite{Clauser69,Clauser74,Garg87}, which is too high to be
realized with current technology. To close the detection loophole,
one should either improve the performances of the devices or
decrease the requirements on the detection efficiency. For Bell test
with CH inequality, no fair-sampling assumption is needed. If Bell
violation is obtained, the lhv theory is excluded. When
non-maximally entangled quantum states are used, it is proven that
one can realize the Bell violation with the detection efficiency
higher than $66.7\%$~\cite{Eberhard93,Brunner07,Cabello07}.

Because comparably lower detection efficiency can be used for Bell
test with CH inequality to realize the Bell violation, it has
attracted wide interesting in optical Bell experiment with present
detection technology~\cite{Giustina13,Christensen13}. In order to
maximize the Bell violation in Bell test with CH inequality,
however, special states and measurement directions should be
chosen~\cite{Eberhard93,Brunner07,Cabello07}. Furthermore, the
tolerable noise is very low when inefficient detectors are used. To
our knowledge, Bell test with all loopholes closed is still the aim
pursued by scientists in the field of fundamental physics.

Recently, measurement-device independent quantum key distribution
(MDI-QKD) protocol has been proposed, where the EPR pairs
post-selected in the Bell basis measurement ensure the security of
the protocol~\cite{Lo12,Biham96,Inamori02}. Entanglement is the
essential nature of quantum physics, it does not matter whether the
entanglement is generated by a parametric-down-conversion (PDC)
source or post-selected in the Bell basis measurement. Thus the the
entanglement generated in the latter way may also be used in the
Bell experiment to test the lhv theory. In this letter, a
time-reversal Bell test protocol is proposed, where EPR pairs
post-selected in the Bell basis measurement are used. Our protocol
can exclude the lhv theory with available detection technology
presently.

\section{Time-reversal Bell test}

In Bell test with CHSH inequality, the lhv theory requires that
\begin{equation}\label{chshinequality}
S_{\rm
CHSH}\equiv\langle{A_{1}B_{1}+A_{1}B_{2}+A_{2}B_{1}-A_{2}B_{2}}\rangle\le2.
\end{equation}
Here $\langle{M}\rangle$ is the expected value of $M$, $A_{i}$,
$B_{j}$ are the observable variables chosen by Alice and by Bob,
with $i,j\in\{1,2\}$. Quantum-mechanically, the maximal value of
$S_{\rm CHSH}$ is proven to be $2\sqrt{2}$~\cite{Tsirelson80}. In
practical condition where the quantum channels are lossy and the
single-photon detectors are inefficient, partial photons are
absorbed within the transmission process and finally missed by the
detectors. If the detection efficiency is $\eta$ at both sides,
closing the detection loophole in Bell test with CHSH inequality
means that~\cite{Garg87}
\begin{equation}\label{detectionloopholefreechsheta}
\frac{2\sqrt{2}\eta^{2}}{\eta^{2}+2\eta(1-\eta)}>2.
\end{equation}
Thus $\eta>82.8\%$ is required to realize detection loophole-free
Bell test with CHSH inequality.

In Bell test with CH inequality, the lhv theory suggests
\begin{equation}\label{chinequality}
S_{\rm
CH}=P(a_{1},b_{1})+P(a_{2},b_{1})+P(a_{1},b_{2})-P(a_{2},b_{2})-P(a_{1})-P(b_{1})\le0,
\end{equation}
where $P(a_{i})$ and $P(b_{j})$ are the probabilities that Alice and
Bob have efficient detections in $a_{i}$ and $b_{j}$, respectively.
The most common bi-partite entanglement states used in the Bell
experiments are the Bell states
\begin{equation}\label{bellstate}
\begin{array}{lll}
|\Phi_{AB}^{+}\rangle&=&\frac{1}{\sqrt{2}}\left(|00\rangle_{AB}+|11\rangle_{AB}\right),\\
|\Phi_{AB}^{-}\rangle&=&\frac{1}{\sqrt{2}}\left(|00\rangle_{AB}-|11\rangle_{AB}\right),\\
|\Psi_{AB}^{+}\rangle&=&\frac{1}{\sqrt{2}}\left(|01\rangle_{AB}+|10\rangle_{AB}\right),\\
|\Psi_{AB}^{-}\rangle&=&\frac{1}{\sqrt{2}}\left(|01\rangle_{AB}-|10\rangle_{AB}\right).
\end{array}
\end{equation}
If inefficient detectors are used, however, the value of the CH
polynomial is maximized by
\begin{equation}\label{nonmaximallyentangledstate}
\rho^{\theta}_{AB}=(1-p)|\Phi^{\theta}_{AB}\rangle\langle\Phi^{\theta}_{AB}|+p\frac{\mathbf{I}}{4},
\end{equation}
where
$|\Phi^{\theta}_{AB}\rangle=\cos\theta|00\rangle_{AB}+\sin\theta|11\rangle_{AB}$,
and $\mathbf{I}$ is the identity matrix used to denote the white
noise. The minimum detection efficiency to violate the CH inequality
can be reduced to be $66.7\%$ by choosing appropriate $\theta$, $p$,
$a_{i}$s and $b_{j}$s~\cite{Eberhard93,Brunner07,Cabello07}.

The MDI-QKD has found great interesting in quantum information as
well as in fundamental physics fields. In the protocol, both Alice
and Bob have well controlled state-preparing devices. They randomly
choose their bases to prepare their quantum states on their photons
which are then distributed to the third party, Eve, who may be
dishonest and harmful to the security of the communication. Eve is
required to implement Bell basis measurement on the incoming photons
and publish her measurement outcomes. The MDI-QKD is proven to be a
time-reversal entanglement-based QKD
protocol~\cite{Lo12,Biham96,Inamori02}. The monogamy of entanglement
can thus ensure the security of the key
bits~\cite{Hill97,Wootters98,Coffman2000,Osborne2006,Horodecki2009}.
As entanglement is post-selected within Bell basis measurement, it
may be used in Bell test to violate the Bell inequalities. The
time-reversal Bell test is described as followings.

\begin{figure}
\centering
\includegraphics[width=0.5\textwidth,trim=0 30 0 30,clip]{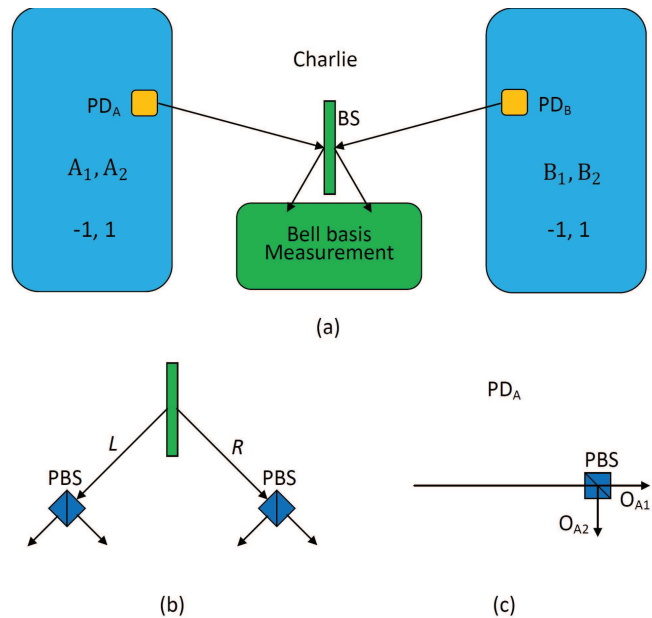}
\caption{(Color online) (a) Alice randomly prepares her photons with
randomly chosen bases, $A_{1}$ and $A_{2}$, and bit values, $-1$ and
$1$. Similarly, Bob randomly chooses his bases between $B_{1}$ and
$B_{2}$, and bit values $-1$ and $1$ to prepare his photons. The
photons are transferred to Charlie who mixes the paths of the
photons with a beam splitter (BS) and implements Bell basis
measurement on them. (b) Scheme of partial Bell basis measurement
with optical systems. It can be used to differentiate
$|\Psi^{+}_{AB}\rangle$ from $|\Psi^{-}_{AB}\rangle$. The former
happens when both Channel $L$ and Channel $R$ have outputs of
different polarizations, while the latter occurs when either Channel
$L$ or Channel $R$ has outputs of different polarizations. (c)
Schematic of $\rm PD_{A}$. Alice departs different polarization
components of her photon with a polarization beam splitter (PBS).}
\label{fig:schematic}
\end{figure}

(a) As is shown in Fig.\ref{fig:schematic} (a), Alice chooses her
basis randomly to be $A_{1}$ and $A_{2}$, and Bob randomly chooses
his bases as $B_{1}$ and $B_{2}$. They also randomly choose their
bit values between $-1$ and $1$. The photons with well prepared
states are transferred to their trust partner, Charlie, who is set
between them.

(b) The photons are guided to coincide on the beam splitter (BS) and
measured by a set of Bell basis measurement device. Charlie
announces Alice and Bob his measurement outcomes. That is, which
quantum state is obtained within the measurement,
$|\Psi^{+}_{AB}\rangle$, $|\Psi^{-}_{AB}\rangle$,
$|\Phi^{+}_{AB}\rangle$, or $|\Phi^{-}_{AB}\rangle$.

(c) Alice and Bob keep their basis choices and bit value choices
when Charlies's measurement outcomes turn out to be
$|\Phi^{+}_{AB}\rangle$ and $|\Psi^{-}_{AB}\rangle$. They use their
basis choices and bit value choices to calculate the CHSH polynomial
in Eq.(\ref{chshinequality}).

The time-reversal version of the present protocol can be
characterized as followings.

(a$^{\prime}$) Charlie is set between Alice and Bob. He randomly
prepares his quantum states to be $|\Psi^{+}_{AB}\rangle$,
$|\Psi^{-}_{AB}\rangle$, $|\Phi^{+}_{AB}\rangle$, and
$|\Phi^{-}_{AB}\rangle$. Then he distributes Photon A to Alice and
Photon B to Bob.

(b$^{\prime}$) Alice takes Photon A and Bob takes Photon B. Alice
randomly chooses her measurement settings to be $A_{1}$ and $A_{2}$,
while Bob randomly selects his measurement settings to be $B_{1}$
and $B_{2}$. They carry out projective measurements on the quantum
states of their photons, respectively.

(c$^{\prime}$) Alice and Bob publish their basis choices and
measurement outcomes. At the same time, Charlie announces Alice and
Bob which Bell state has been generated in his lab. They use the
$|\Phi^{+}_{AB}\rangle$ and $|\Psi^{-}_{AB}\rangle$ events to
calculate the CHSH polynomial.

Thus the present protocol is a time-reversal Bell test protocol.
Quantum principle requires that the events measured to be
$|\Phi^{+}_{AB}\rangle$ make the maximal value of $S_{\rm CHSH}$
reaches $2\sqrt{2}$, while the events measured to be
$|\Psi^{-}_{AB}\rangle$ make the minimal value of $S_{\rm CHSH}$
reaches $-2\sqrt{2}$.

\section{Lhv model of the time-reversal Bell test}

In the present protocol, Alice and Bob prepare their quantum states
in their own labs. They choose their bases and bit values randomly
with genuinely random number generators. If the time intervals
between their state preparations are very short in every run, the
lhv theory does not allow the basis choices and bit value choices at
one side to be affected by those at the other side. Namely,
$\langle{A_{i}B_{j}}\rangle=\langle{A_{i}}\rangle\langle{B_{j}}\rangle$
should be satisfied. One can thus obtain
\begin{equation}\label{lhvofchshinequality}
S_{\rm
CHSH}=\langle{A_{1}}\rangle\langle{B_{1}}\rangle+\langle{A_{1}}\rangle\langle{B_{2}}\rangle+\langle{A_{2}}\rangle\langle{B_{1}}\rangle-\langle{A_{2}}\rangle\langle{B_{2}}\rangle.
\end{equation}
It is easy to verify that $-2\le{S_{\rm CHSH}}\le2$ when
$-1\le\langle{A_{i}}\rangle\le1$ and
$-1\le\langle{B_{j}}\rangle\le1$ are satisfied.

The detection loophole may be the biggest obstacle in excluding the
lhv theory for Bell test with optical systems. In practical Bell
test where inefficient detectors are used, partial photons are
absorbed within the transmission process. Though Bell violation can
be obtained for those events that both sides have efficient
detections, no Bell violation is obtained when the unregistered
events are included. It is because the values on the unregistered
photons are uncertain. This opens the loophole for the
unfair-sampling lhv. If the unfair-sampling lhv exits and decides
the fates of the photon. It can only pick out the maximally violated
events to be registered. If the lhv theory is correct, the events
that both sides have clicks cannot violate the CHSH inequality on
behalf of the whole. The unfair-sampling lhv should also be
considered in the present protocol if inefficient detectors are
used. When partial photons are missed within the transmission
process, Charlie cannot make sure which Bell state should be
assigned to the measurement outcomes. If no Bell violation can be
obtained when these uncertain outcomes are incorporated, the
registered $|\Phi^{+}_{AB}\rangle$ or $|\Psi^{-}_{AB}\rangle$ events
cannot be used to violate the CHSH inequality on behalf of the
whole.

Suppose that the detection efficiency at both sides is $\eta$.
Because Alice and Bob randomly choose their bit values in every
state-preparing basis, $|\Phi^{+}_{AB}\rangle$,
$|\Phi^{+}_{AB}\rangle$, $|\Psi^{+}_{AB}\rangle$ and
$|\Psi^{+}_{AB}\rangle$ happen equiprobably in the Bell basis
measurement. Charlie cannot decide which state is generated in his
Bell basis measurement until both sides have efficient pulses
injecting in his lab and finally detected by his detectors. With
probability $\frac{\eta^{2}}{4}$, Charlie has a register of
$|\Phi^{+}_{AB}\rangle$ in his Bell basis measurement. The
probability Charlie fails to read out the measurement outcomes is
$4\times\frac{(1-\eta^{2})}{4}$. From the time-reversal aspect, the
measurement outcomes $|\Phi^{+}_{AB}\rangle$ mean that the
corresponding states prepared by Alice and by Bob maximize $S_{\rm
CHSH}$ to be $2\sqrt{2}$. Because the four Bell states are
equiprobable, the value of $S_{\rm CHSH}$ for the uncertain
measurement outcomes is averaged to be $0$. To refute the
unfair-sampling lhv, one obtains that
\begin{equation}\label{refuteunfairsamplinglhv}
\frac{2\sqrt{2}\eta^{2}}{\eta^{2}+4(1-\eta^{2})}>2.
\end{equation}
Thus $\eta>95.19\%$ is required to excluding the unfair-sampling
lhv. It is to higher to be realized with present technology.

In normal Bell experiments, Alice and Bob are required to randomly
choose their measurement settings~\cite{Scheidl10}. Or else, the
deterministic lhv may exist~\cite{Selleri98}. Experimentally, faked
Bell violation has been realized perfectly if Alice's and Bob's
freedoms to choose their bases are not allowed~\cite{Gerhardt11}. In
the present protocol, the basis choices and the bit value choices of
Alice's and Bob's states are well chosen as soon as they are
prepared. It means that their basis choices are determined when the
photons are prepared even if Alice and Bob can choose the bases
randomly. Thus deterministic lhv should also be considered in the
present protocol.

\section{Excluding the lhv theory with modified time-reversal Bell test}

The lhv theory is introduced to make the description of quantum
physics complete. If the lhv theory is correct, the lhvs should be
essential attributes of the quantum world. In Bell test with optical
systems, the lhvs are parasitic on the states of the photons. Thus
the lhvs should have the same flying speed as that of the photons.

Suppose that the state encoded on Alice's photon is
$\gamma_{A}|0\rangle_{A}+\delta_{A}|1\rangle_{A}$, and the state
encoded on Bob's photon is
$\gamma_{B}|0\rangle_{B}+\delta_{B}|1\rangle_{B}$. The joint state
on the photons can thus be written as
$\Gamma_{AB}=\gamma_{A}\gamma_{B}|0\rangle_{A}|0\rangle_{B}+\gamma_{A}\delta_{B}|0\rangle_{A}|1\rangle_{B}+\delta_{A}\gamma_{B}|1\rangle_{A}|0\rangle_{B}+\delta_{A}\delta_{B}|1\rangle_{A}|1\rangle_{B}$.
According to the time-reversal formation of entanglement, after the
BS, the part
$\gamma_{A}\gamma_{B}|0\rangle_{A}|0\rangle_{B}+\delta_{A}\delta_{B}|1\rangle_{A}|1\rangle_{B}$
collapses into $|\Phi^{+}_{AB}\rangle$ and $|\Phi^{-}_{AB}\rangle$
within the Bell basis measurement. For the part
$\gamma_{A}\delta_{B}|0\rangle_{A}|1\rangle_{B}+\delta_{A}\gamma_{B}|1\rangle_{A}|0\rangle_{B}$,
however, it collapses into $|\Psi^{+}_{AB}\rangle$ and
$|\Psi^{-}_{AB}\rangle$ in the Bell basis measurement after the BS.

Experimentally, it is still difficult to realize full Bell basis
measurement. As is shown in Fig.\ref{fig:schematic} (b), we consider
the partial Bell basis measurement case where only
$|\Psi^{+}_{AB}\rangle$ and $|\Psi^{-}_{AB}\rangle$ can be
differentiated. With simple calculation, one can obtain
\begin{equation}\label{anticorrelatedpurestate}
\begin{array}{ll}
&\gamma_{A}\delta_{B}|0\rangle_{A}|1\rangle_{B}+\delta_{A}\gamma_{B}|1\rangle_{A}|0\rangle_{B}\\
\rightarrow&\frac{(\delta_{A}\gamma_{B}-\gamma_{A}\delta_{B})}{\sqrt{2}}|\Psi^{-}_{AB}\rangle+\frac{i(\delta_{A}\gamma_{B}+\gamma_{A}\delta_{B})}{\sqrt{2}}|\Psi^{+}_{AB}\rangle.
\end{array}
\end{equation}
The probabilities for $|\Psi^{+}_{AB}\rangle$ and
$|\Psi^{-}_{AB}\rangle$ to occur are different, and their total
probability is calculated to be
$\gamma_{A}^{2}\delta_{B}^{2}+\delta_{A}^{2}\gamma_{B}^{2}$.

It is impossible to filter out the part
$\gamma_{A}\gamma_{B}|0\rangle_{A}|0\rangle_{B}+\delta_{A}\delta_{B}|1\rangle_{A}|1\rangle_{B}$
from the part
$\gamma_{A}\delta_{B}|0\rangle_{A}|1\rangle_{B}+\delta_{A}\gamma_{B}|1\rangle_{A}|0\rangle_{B}$
without destroying $\Gamma_{AB}$. If measurements are carried out on
$\Gamma_{AB}$ in the rectilinear basis, four possible measurement
outcomes occur, namely, $|0\rangle_{A}|0\rangle_{B}$,
$|1\rangle_{A}|1\rangle_{B}$, $|0\rangle_{A}|1\rangle_{B}$, and
$|1\rangle_{A}|0\rangle_{B}$. As is shown in Fig.\ref{fig:schematic}
(c), $\rm PD_{A}$ and $\rm PD_{B}$ are introduced to realize the
measurements. The $\rm PD_{A}$ is composed with a polarization beam
splitter (PBS) and several segments of quantum channels. The photon
with well prepared quantum state is fed into the PBS whose outputs
are then forwarded to Charlie. $\rm PD_{B}$ is operated by Bob in
the same way.

If the bases of both PBSes are aligned with the rectilinear basis,
after the PBSes, the joint state on Photon A and Photon B is
$\gamma_{A}^{2}\gamma_{B}^{2}|00\rangle_{AB}\langle00|+\gamma_{A}^{2}\delta_{B}^{2}|01\rangle_{AB}\langle01|+\delta_{A}^{2}\gamma_{B}^{2}|10\rangle_{AB}\langle10|+\delta_{A}^{2}\delta_{B}^{2}|11\rangle_{AB}\langle11|$.
The part
$\gamma_{A}^{2}\delta_{B}^{2}|01\rangle_{AB}\langle01|+\delta_{A}^{2}\gamma_{B}^{2}|10\rangle_{AB}\langle10|$
is collapsed to be $|\Psi^{+}_{AB}\rangle$ and
$|\Psi^{-}_{AB}\rangle$, while the left is collapsed to be
$|\Phi^{+}_{AB}\rangle$ or $|\Phi^{-}_{AB}\rangle$ within the Bell
basis measurement. Because that
\begin{equation}\label{anticorrelatedmixedstate}
\begin{array}{lll}
|01\rangle_{AB}&\rightarrow&\frac{1}{\sqrt{2}}(i|\Psi^{+}\rangle_{AB}+|\Psi^{-}\rangle_{AB}),\\
|10\rangle_{AB}&\rightarrow&\frac{1}{\sqrt{2}}(i|\Psi^{+}\rangle_{AB}-|\Psi^{-}\rangle_{AB}),
\end{array}
\end{equation}
$|\Psi^{+}\rangle_{AB}$ and $|\Psi^{-}\rangle_{AB}$ happen with
equal probability
$\frac{\gamma_{A}^{2}\delta_{B}^{2}+\delta_{A}^{2}\gamma_{B}^{2}}{2}$.
The total probability of these two events is calculated to be
$\gamma_{A}^{2}\delta_{B}^{2}+\delta_{A}^{2}\gamma_{B}^{2}$, which
is the same as that when no PBS is arranged in the experiment.
Compared (\ref{anticorrelatedmixedstate}) with
(\ref{anticorrelatedpurestate}), though the PBSes destroy the
original quantum states of Alice and Bob, the anti-correlated
components of Alice's and Bob's quantum states are completely picked
out to be measured as $|\Psi^{+}_{AB}\rangle$ or
$|\Psi^{-}_{AB}\rangle$ in both cases. The total probabilities of
$|\Psi^{+}_{AB}\rangle$ and $|\Psi^{-}_{AB}\rangle$ are calculated
to be the same in both cases.

Alice's and Bob's states can be formally rewritten as
$\frac{(\gamma_{A}+\delta_{A})}{\sqrt{2}}|+\rangle_{A}+\frac{(\gamma_{A}-\delta_{A})}{\sqrt{2}}|-\rangle_{A}$,
and
$\frac{(\gamma_{B}+\delta_{B})}{\sqrt{2}}|+\rangle_{B}+\frac{(\gamma_{B}-\delta_{B})}{\sqrt{2}}|-\rangle_{B}$,
where $\{|+\rangle=\frac{1}{\sqrt{2}}(|0\rangle+|1\rangle),
|-\rangle=\frac{1}{\sqrt{2}}(|0\rangle-|1\rangle)\}$ is the diagonal
basis. Accordingly, the joint state on Photon A and Photon B is
$\frac{(\gamma_{A}+\delta_{A})(\gamma_{B}+\delta_{B})}{2}|+\rangle_{A}|+\rangle_{B}+\frac{(\gamma_{A}+\delta_{A})(\gamma_{B}-\delta_{B})}{2}|+\rangle_{A}|-\rangle_{B}+\frac{(\gamma_{A}-\delta_{A})(\gamma_{B}+\delta_{B})}{2}|-\rangle_{A}|+\rangle_{B}+\frac{(\gamma_{A}-\delta_{A})(\gamma_{B}-\delta_{B})}{2}|-\rangle_{A}|-\rangle_{B}$.
After the BS, the anti-correlated components in this joint state is
transformed to be
\begin{equation}\label{transforminthediagonalbasisafterthebs}
\begin{array}{lll}
&
&\frac{(\gamma_{A}+\delta_{A})(\gamma_{B}-\delta_{B})}{2}|+\rangle_{A}|-\rangle_{B}+\frac{(\gamma_{A}-\delta_{A})(\gamma_{B}+\delta_{B})}{2}|-\rangle_{A}|+\rangle_{B}\\
&\rightarrow&\frac{(\delta_{A}\gamma_{B}-\gamma_{A}\delta_{B})}{\sqrt{2}}|\Psi^{-}_{AB}\rangle_{\rm
diagonal}+\frac{i(\gamma_{A}\gamma_{B}-\delta_{A}\delta_{B})}{\sqrt{2}}|\Psi^{+}_{AB}\rangle_{\rm
diagonal}.
\end{array}
\end{equation}
This relation is founded because that
\begin{equation}\label{thebstransformationofdiagonalbasis}
\begin{array}{lll}
|+-\rangle_{AB}&\rightarrow&\frac{1}{\sqrt{2}}(i|\Psi^{+}_{AB}\rangle+|\Psi^{-}_{AB}\rangle)_{\rm diagonal},\\
|-+\rangle_{AB}&\rightarrow&\frac{1}{\sqrt{2}}(i|\Psi^{+}_{AB}\rangle-|\Psi^{-}_{AB}\rangle)_{\rm
diagonal}.
\end{array}
\end{equation}
Here the subscript $\rm diagonal$ means these states are
characterized in the diagonal basis. When the measurement basis is
transformed from the diagonal basis to the rectilinear basis,
$|\Psi^{-}_{AB}\rangle_{\rm diagonal}$ is formally invariant, while
$|\Psi^{+}_{AB}\rangle_{\rm diagonal}$ is converted to be
$|\Phi^{-}_{AB}\rangle$. Thus the probabilities of
$|\Psi^{-}_{AB}\rangle_{\rm diagonal}$ and
$|\Psi^{+}_{AB}\rangle_{\rm diagonal}$ may be different when Alice's
and Bob's state are directly feeded in the Bell basis measurement
device, and their total probability is calculated to be
$\frac{(\gamma_{A}\gamma_{B}-\delta_{A}\delta_{B})^{2}+(\gamma_{A}\delta_{B}-\delta_{A}\gamma_{B})^{2}}{2}$.

When the bases of the PBSes are chosen to be aligned with the
diagonal basis, the joint state on the photons emitting from the
PBSes is
$\frac{(\gamma_{A}+\delta_{A})^{2}(\gamma_{B}+\delta_{B})^{2}}{4}|++\rangle_{AB}\langle++|+\frac{(\gamma_{A}+\delta_{A})^{2}(\gamma_{B}-\delta_{B})^{2}}{4}|+-\rangle_{AB}\langle+-|+\frac{(\gamma_{A}-\delta_{A})^{2}(\gamma_{B}+\delta_{B})^{2}}{4}|-+\rangle_{AB}\langle-+|+\frac{(\gamma_{A}-\delta_{A})^{2}(\gamma_{B}-\delta_{B})^{2}}{4}|--\rangle_{AB}\langle--|$.
The anti-correlated components of the state is converted to
$|\Psi^{+}_{AB}\rangle_{\rm diagonal}$ and
$|\Psi^{-}_{AB}\rangle_{\rm diagonal}$ in the Bell basis
measurement. According to
(\ref{thebstransformationofdiagonalbasis}), both
$|\Psi^{+}_{AB}\rangle_{\rm diagonal}$ and
$|\Psi^{-}_{AB}\rangle_{\rm diagonal}$ happen with probability
$\frac{(\gamma_{A}\gamma_{B}-\delta_{A}\delta_{B})^{2}+(\gamma_{A}\delta_{B}-\delta_{A}\gamma_{B})^{2}}{4}$.
Thus their total probability is
$\frac{(\gamma_{A}\gamma_{B}-\delta_{A}\delta_{B})^{2}+(\gamma_{A}\delta_{B}-\delta_{A}\gamma_{B})^{2}}{2}$,
which is equal to that when no PBS is set in Alice's and Bob's labs.
It means the PBSes in the diagonal basis can also pick out the
anti-correlated events to be measured as $|\Psi^{+}_{AB}\rangle_{\rm
diagonal}$ and $|\Psi^{-}_{AB}\rangle_{\rm diagonal}$ in the Bell
basis measurement.

Experimentally, Alice's and Bob's bit value choices in every basis
are randomly chosen. Namely,
$\{\gamma_{A}|0\rangle_{A}+\delta_{A}|1\rangle_{A},
-\delta_{A}|0\rangle_{A}+\gamma_{A}|1\rangle_{A}\}$ are equiprobably
prepared by Alice, and
$\{\gamma_{B}|0\rangle_{B}+\delta_{B}|1\rangle_{B},
-\delta_{B}|0\rangle_{B}+\gamma_{B}|1\rangle_{B}\}$ are equiprobably
prepared by Bob. Thus Both Alice's and Bob's states are
characterized as identity matrix, and their joint state is an
identity matrix in the $4$-dimensional Hilbert space.

A modified time-reversal Bell test with the lhv theory excluded can
be realized with the following steps.

(A) Alice has two copies of $\rm PD_{A}$ ($\rm PD^{1}_{A}$, and $\rm
PD^{2}_{A}$) , and Bob has two copies of $\rm PD_{B}$ ($\rm
PD^{1}_{B}$ and $\rm PD^{2}_{B}$). Alice's bases are randomly chosen
to be $A_{1}=\sin\theta\hat{\sigma}_{x}+\cos\theta\hat{\sigma}_{z}$,
$A_{2}=\sin(\theta+\frac{\pi}{2})\hat{\sigma}_{x}+\cos(\theta+\frac{\pi}{2})\hat{\sigma}_{z}$.
Bob's bases are randomly chosen to be
$B_{1}=\sin(\theta+\frac{\pi}{4})\hat{\sigma}_{x}+\cos(\theta+\frac{\pi}{4})\hat{\sigma}_{z}$,
$B_{2}=\sin(\theta-\frac{\pi}{4})\hat{\sigma}_{x}+\cos(\theta-\frac{\pi}{4})\hat{\sigma}_{z}$.
Their bit values are randomly chosen as $-1$ and $1$.

(B) According to their basis choices and bit value choices, Alice
and Bob prepare their states in the labs. Every state prepared by
Alice has two copies. One is for $\rm PD^{1}_{A}$ and the other is
for $\rm PD^{2}_{A}$. Similarly, every state prepared by Bob also
has two copies. One is for $\rm PD^{1}_{B}$ and the other is for
$\rm PD^{2}_{B}$. The bases of the PBSes in $\rm PD^{1}_{A}$ and
$\rm PD^{1}_{B}$ are aligned with the rectilinear basis, while the
those of the PBSes in $\rm PD^{2}_{A}$ and $\rm PD^{2}_{B}$ are
aligned with the diagonal basis.

(C) The photons with the well prepared states are fed into the
PBSes. Alice sets her single-photon-detectors (SPDs) at every output
of her PBSes, and so does Bob. They record all the registers on
their detectors.

(D) Alice and Bob announce their measurement outcomes through the
public channels. The events are kept when both $\rm PD^{1}_{A}$ and
$\rm PD^{1}_{B}$ emit anti-correlated clicks, and simultaneously,
both $\rm PD^{2}_{A}$ and $\rm PD^{2}_{B}$ also generate
anti-correlated detections.

(E) The experiment goes on until Alice and Bob keep enough
experimental data. They use their bit values and basis choices of
the kept events to calculate the value of $S_{\rm CHSH}$.

In step (C), Alice and Bob measure on the photons emitted out from
their PBSes rather than forward them to Charlie. If the time
intervals between Alice's and Bob's measurements are very short, the
lhv at Alice's side cannot reach Bob's lab before Bob's photon is
clicked on his detectors, and vice versa. When the Bell violation is
obtained, from Eq.(\ref{lhvofchshinequality}), it cannot be
contributed by the lhvs. The anti-correlated results from $\rm
PD^{1}_{A}$ and $\rm PD^{1}_{B}$ means that the measurement outcomes
should be $|\Psi^{+}_{AB}\rangle$ or $|\Psi^{-}_{AB}\rangle$ if
Photon A and Photon B passed the BS and were measured in the Bell
basis. When Alice and Bob have anti-correlated results in $\rm
PD^{2}_{A}$ and $\rm PD^{2}_{B}$, however, Charlie's measurement
outcome should be $|\Phi^{-}_{AB}\rangle$ or $|\Psi^{-}_{AB}\rangle$
if the photons were not measured directly. Thus when the conditions
in step (D) are satisfied, the measurement outcome of Charlie should
be $|\Psi^{-}_{AB}\rangle$ if the photons at both sides passed the
BS and were measured in the Bell basis.

For $\rm PD^{1}_{A}$ and $\rm PD^{1}_{B}$, the state on their
anti-correlated outputs is $\rho^{\rm rectilinear}_{\rm
anti-correlated}=\frac{1}{2}(|01\rangle_{AB}\langle01|+|10\rangle_{AB}\langle10|)$.
For $\rm PD^{2}_{A}$ and $\rm PD^{2}_{B}$, however, the state on the
anti-correlated outputs is $\rho^{\rm diagonal}_{\rm
anti-correlated}=\frac{1}{2}(|+-\rangle_{AB}\langle+-|+|-+\rangle_{AB}\langle-+|)$.
It is easy to verify that $\rm{Tr}(\rho^{\rm rectilinear}_{\rm
anti-correlated}\hat{\sigma}_{z}^{A}\otimes\hat{\sigma}_{z}^{B})=-1$,
$\rm{Tr}(\rho^{\rm digonal}_{\rm
anti-correlated}\hat{\sigma}_{x}^{A}\otimes\hat{\sigma}_{x}^{B})=-1$,
and $\rm{Tr}(\rho^{\rm s}_{\rm
anti-correlated}\hat{\sigma}_{z}^{A}\otimes\hat{\sigma}_{x}^{B})=\rm{Tr}(\rho^{\rm
s}_{\rm
anti-correlated}\hat{\sigma}_{x}^{A}\otimes\hat{\sigma}_{z}^{B})=0$,
with $s\in\{\rm rectilinear, diagonal\}$. If there is a single
state, $\rho_{\rm anti-correlated}$, who combines the characters of
both $\rho^{\rm rectilinear}_{\rm anti-correlated}$ and $\rho^{\rm
diagonal}_{\rm anti-correlated}$, it can be proven to be $\rho_{\rm
anti-correlated}=|\Psi^{-}_{AB}\rangle\langle\Psi^{-}_{AB}|$, and
$S_{\rm CHSH}^{|\Psi^{-}_{AB}\rangle}=-2\sqrt{2}$. It means that if
the conditions of step (D) in the modified time-reversal Bell test
are satisfied, Alice and Bob succeed in selecting $S_{\rm
CHSH}^{|\Psi^{-}_{AB}\rangle}$. From the aspect of time-reversal
formation of entanglement, the CHSH inequality is violated and the
lhv theory is excluded in the the modified time-reversal Bell test.
From (\ref{anticorrelatedpurestate}) and
(\ref{transforminthediagonalbasisafterthebs}), the experiment
succeeds with probability
$\frac{(\gamma_{A}^{2}\delta_{B}^{2}-\delta_{A}^{2}\gamma_{B}^{2})^{2}}{4}$.

\section{Discussion and conclusion}

In this letter, we have raised the time-reversal Bell test protocol.
Accordingly, we have proposed the lhv model for it. Analyses were
given on the unfair-sampling lhv and the deterministic lhv which may
contribute the Bell violation in the lhv theory. By simply measuring
their own prepared states directly in the rectilinear basis or the
diagonal basis, Alice and Bob are able to differentiate the
anti-correlated components from the correlated components of the
states destructively. In the rectilinear basis, the anti-correlated
components contribute to $|\Psi^{+}_{AB}\rangle$ and
$|\Psi^{-}_{AB}\rangle$. In the diagonal basis, the anti-correlated
contribute to $|\Phi^{-}_{AB}\rangle$ and $|\Psi^{-}_{AB}\rangle$.
In every run, Alice and Bob each prepares two copies of their
states. The anti-correlated components of one copy of their joint
state are distinguished in the rectilinear basis, and those of the
other copy are distinguished in the diagonal basis. The overlapped
anti-correlated results in their detectors should contribute to
$|\Psi^{-}_{AB}\rangle$ if they passed the BS and were measured in
the Bell basis. As Alice and Bob detect their states in their own
labs, the lhvs are prevented at both sides to contact with each
other. Thus the violation of the CHSH inequality supports the
correctness of quantum principles and excludes the lhv theory. Our
protocol only picks up the efficient detections of Alice and Bob
locally, it does not require high detection efficiency. Thus it can
be realized with present technology.

\section*{Acknowledgement}

This work is supported by MOST 2013CB922003 of the National Key
Basic Research Program of China, NSFC under Grant No. 61378011,
Program for Science and Technology Innovation Talents in
Universities of Henan Province (Grant No. 2012HASTIT028) and program
for Science and Technology Innovation Research Team in University of
Henan Province (Grant No. 13IRTSTHN020).

\section*{Author contribution statement}

Y.T wrote the manuscript text, drew Fig. 1, and reviewed the
manuscript.

\section*{Competing financial interests}
The authors declare no competing financial interests.

\end{document}